\documentclass[aps,prd, twocolumn,superscriptaddress,nofootinbib,preprintnumbers]{revtex4-1}
\usepackage{mathrsfs}
\usepackage{amsfonts}
\usepackage{amsmath}
\usepackage{array}
\usepackage{verbatim}
\usepackage{epsfig}
\usepackage{graphicx}
\usepackage{color}

\newcommand{\beq}{\begin{equation}}
\newcommand{\eeq}{\end{equation}}
\newcommand{\bea}{\begin{eqnarray}}
\newcommand{\eea}{\end{eqnarray}}
\newcommand{\nn}{\nonumber}

\begin{document}
\preprint{
{\vbox {
\hbox{\bf MSUHEP-18-020}
}}}
\vspace*{0.2cm}

\title{Transverse Momentum Resummation for $s$-channel\\ single top quark production at the LHC}

\author{Peng Sun}
\email{pengsun@msu.edu}
\affiliation{Department of Physics and Institute of Theoretical Physics, Nanjing Normal University, Nanjing, Jiangsu, 210023, China}

\author{Bin Yan}
\email{yanbin1@msu.edu}
\affiliation{Department of Physics and Astronomy,
\\Michigan State University, East Lansing, MI 48824, USA}

\author{C.-P. Yuan}
\email{yuan@pa.msu.edu}
\affiliation{Department of Physics and Astronomy,
\\Michigan State University, East Lansing, MI 48824, USA}

\begin{abstract}

We study the soft gluon radiation effects for the $s$-channel single top quark production at the LHC.
By applying the transverse momentum
dependent factorization formalism, the large logarithms about the small total transverse momentum ($q_\perp$)
of the single-top plus one-jet final state system, are resummed
to all orders in the expansion of the strong interaction coupling at the accuracy of Next-to-Leading Logarithm (NLL).
We compare our numerical results with PYTHIA and find  that both the $q_\perp$ and $\phi^*$ observables 
from PYTHIA  are consistent with our prediction. Furthermore,  we 
point out the soft gluon radiation effects from the final state become significant in this process, especially for the boosted kinematical region.

\end{abstract}

\maketitle

\section{Introduction}
Single top quark production is an important source of top quarks at the Large Hadron Collider (LHC). There are three production modes, $s$-channel, $t$-channel and $tW$ associated production.  In addition to measuring the $V_{tb}$ Cabibbo-Kobayashi-Maskawa (CKM) matrix element~\cite{Berger:2009hi,Cao:2015qta}, they  are also sensitive to different kinds of  new physics (NP) models beyond the standard model (SM)~\cite{Dawson:1984gx,Willenbrock:1986cr,Dawson:1986tc,Yuan:1989tc,Ladinsky:1992vv,Carlson:1993dt,Carlson:1995ck,Tait:1996dv,Li:1996ir,Li:1997qf,Tait:1997fe,He:1998ie,Tait:2000sh,Malkawi:1996fs,Hsieh:2010zr,Cao:2012ng,Drueke:2014pla,Cao:2013ud,Cao:2006wk,Berger:2011hn,Berger:2011xk,Kane:1991bg,Carlson:1994bg,Chen:2005vr,Cao:2007ea,Berger:2009hi,Fabbrichesi:2014wva,Bernardo:2014vha,Cao:2015doa,Prasath:2014mfa,Hioki:2015env,Zhang:2016omx,
Birman:2016jhg,Boos:2016zmp,Berger:2009hi,Cao:2015qta},
such as  new heavy gauge boson $W^\prime$~\cite{Malkawi:1996fs,Hsieh:2010zr,Cao:2012ng},   fermions~\cite{Cao:2006wk,Berger:2011hn,Berger:2011xk}, scalars~\cite{Drueke:2014pla,Cao:2013ud}, and $Wtb$ anomalous couplings~\cite{Kane:1991bg,Carlson:1994bg,Chen:2005vr,Cao:2007ea,AguilarSaavedra:2008gt,Berger:2009hi,Rindani:2011pk,Fabbrichesi:2014wva,Bernardo:2014vha,Cao:2015doa,Prasath:2014mfa,Hioki:2015env,Zhang:2016omx,
Birman:2016jhg,Boos:2016zmp,Jueid:2018wnj}.  Compared with the $t$-channel and $tW$ associated production, the $s$-channel single top quark event is more sensitive to NP effects induced by heavy resonance states.  Therefore, precisely study the single top quark production processes at the LHC become a vital task to test the SM and to search for  new heavy particles. Recently,  both the ATLAS and CMS collaborations have conducted search for  new particles  through $s$-channel single top quark event at the 13 TeV LHC, and concluded that their masses should be larger than about TeV scale~\cite{Sirunyan:2017vkm,Aaboud:2018juj}.  

To further test the SM and search for  NP through the single top quark processes, we should improve the accuracy of the theoretical prediction on its cross section and kinematical distributions. The next-to-leading-order (NLO) QCD correction to the single top quark production has been widely discussed in the literatures~\cite{Zhu:2001hw,Harris:2002md,Cao:2004ky, Cao:2004ap,Campbell:2004ch,Cao:2005pq, Campbell:2005bb,Cao:2008af,Heim:2009ku,Campbell:2009ss,Schwienhorst:2010je,Falgari:2010sf, Frixione:2005vw,Alioli:2009je,Frederix:2012dh,Frederix:2016rdc}.  The dominant part of the next-to-next-to-leading-order (NNLO) QCD corrections to predicting the detailed kinematical distributions, including proper spin correction, in $s$-channel and $t$-channel single top events, have also been discussed in Refs.~\cite{Brucherseifer:2014ama,Berger:2016oht,Berger:2017zof,Liu:2018gxa}.
To go beyond the fixed-order calculations,  the threshold resummation technique is also widely discussed to improve
the prediction on the single-top inclusive production rate~\cite{Kidonakis:2006bu,Kidonakis:2007ej,Kidonakis:2011wy,Kidonakis:2013yoa,Kidonakis:2015wva,Zhu:2010mr,Wang:2010ue}. The accuracy has reached to the next-to-leading-logarithm (NLL) and next-to-next-to-leading-logarithm (NNLL). Recently, the transverse momentum resummation formalism was  proposed in Ref.~\cite{Cao:2018ntd} to improve the kinematical distributions of $t$-channel single top events.  It shows that the sub-leading logarithms  from the  color correlation  between the initial  and final states  play an important role when the final state jet is required to be in the forward region, where the resummation prediction is noticeably different from the PYTHIA parton shower results. Motivated by this, it is important to check on
the kinematical distributions of $s$-channel single top quark events predicted by PYTHIA.

In this work, we apply the  transverse momentum dependent (TMD) resummation technique to study the kinematical distribution of $s$-channel single top quark events, 
\bea
p+p\to W^{\pm\star}\to t (\bar{t})+jet+X.
\eea
The large  logarithms $\ln(Q^2/q_\perp^2)$, with $Q\gg q_\perp$ have been resummed to NLL accuracy, where $Q$ and $q_\perp$ are the invariant mass and the total transverse momentum  of the top quark and jet system, respectively.  The TMD resummation framework has been widely discussed  in the color singlet processes~\cite{Collins:1981uk,Collins:1981va,Collins:1984kg}.  For the processes with more complex color structures, like the heavy colored particle production was  discussed in Refs.~\cite{Zhu:2012ts,Li:2013mia,Zhu:2013yxa}.  Recently, the 
TMD resummation formalism has been extended to discuss  processes involving multijets in the final state; e.g. dijet production~\cite{Sun:2014gfa,Sun:2015doa}, Higgs plus one and two jets production~\cite{Sun:2014lna,Sun:2016kkh,Sun:2016mas,Sun:2018beb}, $Z$ boson and jet associated production~\cite{Sun:2018icb} and $t$-channel single top quark production~\cite{Cao:2018ntd}.  
The soft gluon radiation from the final state will generate  additional large  logarithm $\ln(Q^2/q_\perp^2)$ when  gluons are radiated outside the observed jet cone.  Such logarithms can be resummed under the modified TMD resummation formalism.
As to be shown below, the location and height of the Sudakov peak, in the $q_{\perp}$ distribution of $s$-channel single top quark events, strongly depends on the final state soft gluon radiation. Its effects could be enhanced largely when we focus on the boosted kinematical region of the final state. In contrast to the findings in the  $t$-channel single-top production process, we find that  the resummation calculation in the $s$-channel single-top process  agrees well with PYTHIA prediction.

\section{TMD factorization}
The differential cross section for $pp\to W^{\pm\star}\to t (\bar{t})+jet+X$ can be written as,
\begin{align}
&\frac{d^4\sigma}
{dy_t dy_J d P_{J\perp}^2
d^2q_{\perp}}=\sum_{ab}\nn\\
&\left[\int\frac{d^2\vec{b}}{(2\pi)^2}
e^{-i\vec{q}_\perp\cdot
\vec{b}}W_{ab\to t J}(x_1,x_2,\textbf{b})+Y_{ab\to tJ}\right] \ ,\label{resumy}
\end{align}
where $y_t$ and $y_J$ denote rapidity for the top quark and the  jet, respectively;  $P_{J\perp}$ and $q_{\perp}$ are the transverse momenta  of the jet and  total transverse momentum of the top quark and the jet system, i.e. $\vec{q}_\perp=\vec{P}_{t\perp}+\vec{P}_{J\perp}$, respectively. The $W_{ab\to tJ}$ term contains all order resummation  and $Y_{ab\to tJ}$ term accounts for the difference between the expansion of resummation part and the fixed order corrections, and $x_1, x_2$ are momentum fractions of the incoming hadrons carried by the partons,
\begin{equation}
x_{1,2}=\frac{\sqrt{m_t^2+P^2_{t\perp}}e^{\pm y_t}+\sqrt{P^2_{J\perp}}e^{\pm y_J}}{\sqrt{S}},
\end{equation}
where $m_t$ and $S$ are the top quark mass and the squared collider energy, respectively.

The all order resummation result for $W$-piece  can be written as,
\begin{align}
W_{ab\to tJ}\left(x_1,x_2,\textbf{b}\right)&=x_1\,f_a(x_1,\mu_F=b_0/b_*)
x_2\, f_b(x_2,\mu_F=b_0/b_*) \nn\\
&\times e^{-S_{\rm Sud}(Q^2,\mu_{\rm Res},b_*)}e^{-\mathcal{F}_{NP}(Q^2,\textbf{b})} \nn\\
&\times H_{ab\to tJ}(\mu_{\rm Res},\mu_{\rm Ren})S_{ab\to tJ}(b_0/b_*),
\label{resum}
\end{align}
where $Q^2=\hat{s}=x_1x_2S$, $b_0=2e^{-\gamma_E}$, with $\gamma_E$ being the Euler constant, $f_{a,b}(x,\mu_F)$ are parton distribution functions (PDFs) for the incoming partons $a$ and $b$, 
and $\mu_{\rm Res}$ and $\mu_{\rm Ren}$ represent the resummation and renormalization scales respectively in this process.
Here, $b_*=\textbf{b}/\sqrt{1+\textbf{b}^2/b_{\rm{max}}^2}$ with $b_{\rm {max}}=1.5~{\rm GeV}^{-1}$,
which is introduced to factor out the non-perturbative contribution
$e^{-\mathcal{F}_{NP}(Q^2,b)}$, arising from the large $\textbf{b}$ region (with $\textbf{b} \gg b_*$)
~\cite{Landry:1999an,Landry:2002ix,Sun:2012vc,Su:2014wpa},
\bea
\mathcal{F}_{NP}(Q^2,\textbf{b})=g_1\textbf{b}^2+g_2\ln\dfrac{Q}{Q_0}\ln\dfrac{\textbf{b}}{b_*},\\ \nn
\eea
where $g_1=0.21$, $g_2=0.84$ and $Q_0^2=2.4~{\rm GeV}^2$~\cite{Su:2014wpa}.
$H_{ab\to tJ}$ and $S_{ab\to tJ}$ are the hard and soft factors for this process. Similar to the t-channel single-top production process, 
there are two orthogonal color configurations in the s-channel single-top production process and the resummation calculation should be carried out 
in the color space with matrix form~\cite{Zhu:2010mr,Cao:2018ntd}. 
However, since the color-octet component in this process is much smaller than the color singlet component, we shall only include the color singlet component in our calculation as to be shown below.  By applying the Catani-De Florian-Grazzini (CFG) scheme~\cite{Catani:2000vq} and the TMD factorization in the Collins 2011 scheme~\cite{Collins:2011zzd}, we obtain the hard factor $H_{ab\to tJ}$, at the NLO level, 
\begin{widetext}
\begin{align}
H^{(1)}_{ab\to tJ}&=\dfrac{\alpha_s(\mu_{\rm Ren})}{2\pi}C_FH^{(0)}\left[-\ln^2(\lambda-1)-\dfrac{\ln(\lambda-1)}{\lambda}-2\ln(\lambda-1)
-2\ln(\lambda-1)\ln\dfrac{\hat{s}}{m_t^2}+\ln\dfrac{\mu_{\rm Res}^2}{\hat{s}}
\left(-2\ln(\lambda-1)-\ln\dfrac{\hat{s}}{m_t^2}-\dfrac{11}{2}
\right)\right. \nn\\
&\left.-\dfrac{1}{2}\ln^2\dfrac{\hat{s}}{m_t^2}-\dfrac{5}{2}\ln\dfrac{\hat{s}}{m_t^2}-\dfrac{3}{2}\ln\dfrac{P_{J\perp}^2R^2}{\mu_{\rm Res}^2}+2\mathrm{Li}_2(\lambda)+\dfrac{1}{2}\ln^2\dfrac{P_{J\perp}^2R^2}{\mu_{\rm Res}^2}
-\dfrac{3}{2}\ln^2\dfrac{\mu_{\rm Res}^2}{\hat{s}}+\dfrac{4\pi^2}{3}-\dfrac{15}{2}
\right]
+\delta H^{(1)},
\end{align}
\end{widetext}
where $\lambda=\hat{s}/(\hat{s}-m_t^2)$,  $R$ denotes the jet cone size of the final state jet. Both the loop correction and jet function have been included in the above hard factor.  For the jet function calculation, the dimensional regularization and anti-$k_T$  jet algorithm are adopted in our calculation~\cite{Mukherjee:2012uz,Sun:2015doa},  and an off-shell mass is assigned to the light jet to regulate the light cone singularity in the soft factor calculation. The different treatment of the jet part in the jet function and the soft factor leads to a finite contribution in the hard factor, which does not depending on the jet size. Numerically, it is found to be approximately $\frac{\alpha_s}{2\pi}C_F\frac{\pi^2}{6}$ for quark jet~\cite{Sun:2016kkh}.
This additional factor has been considered as part of  $H^{(1)}$.   The leading order hard matrix element is,
\begin{align}
H^{(0)}(ij\to t\bar{b})&=\dfrac{g^4\hat{t}(\hat{t}-m_t^2)}{4(\hat{s}-m_W^2)^2}|V_{ij}|^2|V_{tb}|^2,
\end{align}
where  $g$ is $SU(2)_L$ gauge coupling. The CKM matrix element $V_{ij}$ needs to change for the corresponding incoming partons, and $m_W$ is $W$-boson mass.  The $\delta H^{(1)}$ term is not proportional to the leading order matrix element,
\begin{equation}
\delta H^{(1)}=\dfrac{\alpha_s}{2\pi}\dfrac{1}{4}\dfrac{g^4C_Fm_t^2}{(\hat{s}-m_W^2)^2}\dfrac{\hat{t}\hat{u}}{\hat{s}}\ln\dfrac{m_t^2}{\hat{s}-m_t^2}|V_{ud}|^2|V_{tb}|^2,
\end{equation}
where $\hat{t}=(p_u-p_{\bar{b}})^2$ and $\hat{u}=(p_{\bar{d}}-p_{\bar{b}})^2$.

The Sudakov form factor ${\cal S}_{\rm Sud}$ resums the leading double logarithm and the sub-leading logarithms,
\begin{align}
S_{\rm Sud}(Q^2,\mu_{\rm Res},b_*)=\int^{\mu_{\rm Res}^2}_{b_0^2/b_*^2}\frac{d\mu^2}{\mu^2}
\left[\ln\left(\frac{Q^2}{\mu^2}\right)A+B_1+B_2 \right.\nn\\
\left.+D_1\ln\frac{Q^2-m_t^2}{P_{J\perp}^2R^2}+
D_2\ln\frac{Q^2-m_t^2}{m_t^2}\right]\ , \label{su}
\end{align}
where  the parameters $A$, $B_1$, $B_2$, $D_1$ and $D_2$ can be expanded perturbatively in $\alpha_s$. At one-loop order,
\begin{align}
&A=C_F\dfrac{\alpha_s}{\pi},\quad & B_1&=-C_F\dfrac{3\alpha_s}{2\pi},\nn\\
&B_2=-C_F\dfrac{\alpha_s}{2\pi}, \quad & D_1 &=D_2=C_F\dfrac{\alpha_s}{2\pi},
\label{su_one}
\end{align}
with $C_F=4/3$ in QCD interaction.
In our numerical calculation, we will also include the $A^{(2)}$ contribution since it is associated with the incoming parton distributions and universal for all processes~\cite{Catani:2000vq}. 
The coefficients $A$ and $B_1$ come from the energy evolution effect in the TMD PDFs~\cite{Ji:2004wu}, so that they only depend on the flavor of the incoming partons of the leading order  scattering processes. The coefficient $B_2$ describes the soft gluon emission from the final state  top quark. The factor $D_1$ and $D_2$ quantifies the effect of soft gluon radiation between the top quark and jet in the final state.

\begin{figure}
\includegraphics[width=0.23\textwidth]{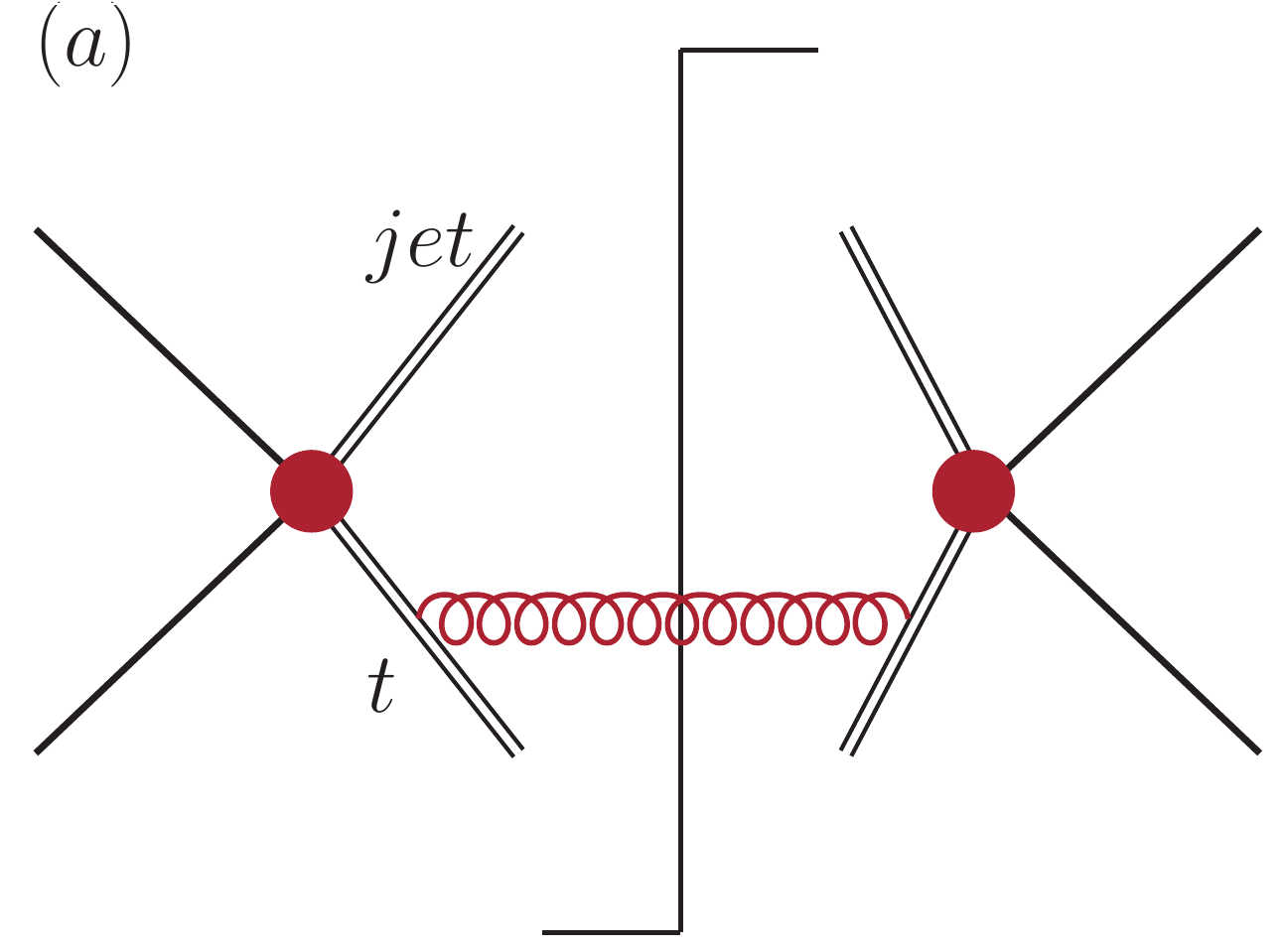}
\includegraphics[width=0.23\textwidth]{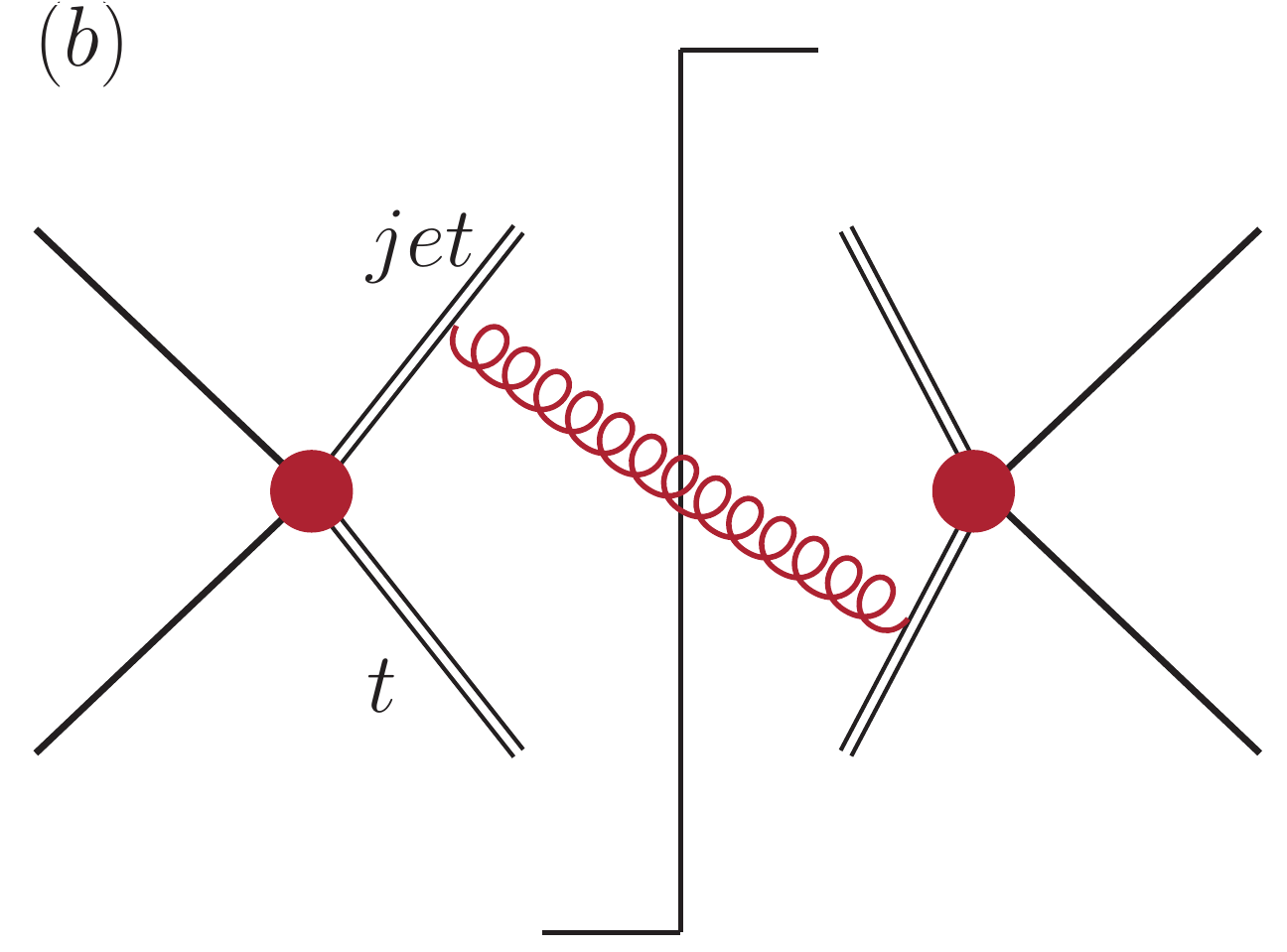}
\caption{Feynman diagrams contributing to the soft functions at the NLO.}
\label{fig:soft}
\end{figure}

The soft function $S_{ab\to tJ}(\mu)$ at the scale $\mu$ 
can be calculated based on the method in Ref.~\cite{Sun:2015doa}.
At one loop order, it is
\begin{align}
S^{(1)}_{ab\to 
tJ}(\mu)=&-\dfrac{\alpha_s}{2\pi}C_F\left[-1+\ln\dfrac{Q^2-m_t^2}{m_t^2}+\ln\dfrac{Q^2-m_t^2}{P_{J\perp^2}R^2}\right]\ln\left(\dfrac{b_*^2}{b_0^2}\mu^2\right)\nn\\
&-\dfrac{\alpha_s}{2\pi}C_F\left[\ln\dfrac{m_t^2}{m_t^2+P_{J\perp}^2}+I_{34}\right].
\end{align}
Hence, $S^{(1)}_{ab\to tJ}(b_0/b_*)$ in Eq.~(\ref{resum}),
evaluated at $\mu=b_0/b_*$, is
\bea
S^{(1)}_{ab\to 
tJ}(b_0/b_*)=-\dfrac{\alpha_s}{2\pi}C_F\left[\ln\dfrac{m_t^2}{m_t^2+P_{J\perp}^2}+I_{34}\right],
\eea
where the first term originates from the contribution of the final state top quark line, as shown in Fig.~\ref{fig:soft}(a), and $I_{34}$ represents the contribution of soft gluon radiation between the final state jet and top quark lines as shown in Fig.~\ref{fig:soft}(b).
In  the narrow jet approximation; i.e. $P_{J\perp}R\to 0$, $I_{34}$ can be written as
\begin{widetext}
\begin{align}
I_{34}&=-\mathrm{Li}_2\dfrac{m_t^2+\hat{t}-\hat{u}}{\hat{t}}-\mathrm{Li}_2\dfrac{(2m_t^2-\hat{s})(m_t^2-\hat{t})}{\hat{s}\hat{t}}+\mathrm{Li}_2\dfrac{(\hat{s}-2m_t^2)\hat{t}}{\hat{s}\hat{u}}-\ln\dfrac{m_t^2-\hat{u}}{m_t^2+\hat{t}-\hat{u}}\ln\dfrac{-m_t^2(m_t^2+\hat{t}-\hat{u})}{\hat{s}\hat{u}}\nn\\
&+\ln\dfrac{-\hat{t}}{m_t^2+\hat{t}-\hat{u}}\ln\dfrac{(m_t^2-\hat{s})(m_t^2+\hat{t}-\hat{u})}{\hat{s}\hat{u}}
+(\hat{t}\leftrightarrow\hat{u})-\ln\dfrac{\hat{s}-m_t^2}{m_t^2}\ln\dfrac{\hat{t}\hat{u}}{m_t^4-(\hat{t}-\hat{u})^2}\nn\\
&-\ln\dfrac{P_{J\perp}^2R^2\hat{s}}{\hat{t}\hat{u}}\ln\dfrac{\hat{s}-m_t^2}{-P_{J\perp}^2R^2}-\dfrac{1}{2}\ln^2\dfrac{P_{J\perp}^2R^2}{\hat{s}-2m_t^2}
-\dfrac{1}{2}\ln^2\dfrac{m_t^2}{2m_t^2-\hat{s}}+\dfrac{1}{2}\ln^2\dfrac{\hat{s}-m_t^2}{2m_t^2-\hat{s}}-\ln\dfrac{\hat{s}-m_t^2}{2m_t^2-\hat{s}}\ln\dfrac{P_{J\perp}^2R^2}{\hat{s}-2m_t^2}\nn\\
&+2\ln\dfrac{\hat{
s}-m_t^2}{2m_t^2-\hat{s}}\ln\dfrac{P_{J\perp}^2R^2}{\hat{s}-m_t^2}+\ln\dfrac{m_t^2}{2m_t^2-\hat{s}}\ln\dfrac{m_t^2\hat{s}}{\hat{t}\hat{u}}-2\ln\dfrac{2m_t^2-\hat{s}}{m_t^2-\hat{s}}\ln\dfrac{m_t^2}{2m_t^2-\hat{s}}-2\mathrm{Li}_2\dfrac{m_t^2}{\hat{s}-m_t^2}-\dfrac{\pi^2}{3}+\mathcal{O}(\cdots),
\end{align}
\end{widetext}
where the $(\cdots)$ term contains contributions proportional to $P_{J\perp}R$, and will be included in the following numerical calculation.

We should note that the non-global logarithms (NGLs) could also contribute to this process.
The NGLs arise from some special kinematics of two soft gluon radiations, in which the first one is radiated outside of the jet which subsequently radiates a second gluon into the jet~\cite{Dasgupta:2001sh,Dasgupta:2002bw,Banfi:2003jj,Forshaw:2006fk}. Numerically, the NGLs are negligible in this process since it starts at $\mathcal{O}(\alpha_s^2)$~\cite{Sun:temp}. Therefore  we will ignore their contributions in the following phenomenology discussion.

\section{Phenomenology of single top quark production}
Before  presenting the result of  resummation effects on the kinematical distributions of the $s$-channel single top quark events, it is important to cross-check the total cross section with the fixed order calculation. In the resummation framework, the NLO total cross section can be divided into two parts, the small $q_{\perp}$ region, which can be obtained by integrating the distribution of the asymptotic part and virtual diagram contribution,  and the large $q_{\perp}$ part, which is infrared safe and can be numerically calculated directly~\cite{Balazs:1997xd}. Thus, the NLO total cross section is given by
\bea
\sigma_{NLO}=\int_0^{q_{\perp,0}^2}dq_{\perp}^2\dfrac{d\sigma_{NLO}^{virtual+real}}{dq_{\perp}^2}+\int_{q_{\perp,0}^2}^{\infty}dq_{\perp}^2\dfrac{d\sigma_{NLO}^{real}}{dq_{\perp}^2},\nn\\
\label{eq:cs}
\eea
where $q_{\perp,0}=1~ {\rm GeV}$ labels the cutoff of $q_{\perp}$. 
In the above equation, the integrand of the first term was obtained by expanding the contribution from the 
$W$-term, cf. Eq.~\ref{resum}, up to order $\alpha_s$, but without including the $Y$-term contribution which 
is small for $q_\perp < 1$ GeV. 
The numerical result of Eq.~\ref{eq:cs}  is found to be slightly different 
from the prediction of MCFM with $\mu_{\rm Ren}=\mu_F=m_t$~\cite{Campbell:2015qma}, ranging from 1.8\% for  $R=0.4$ to 0.3\% for $R=0.2$. Clearly,  this discrepancy arises from the narrow jet approximation we made in our calculation. Following the procedure of Ref.~\cite{Sun:2016kkh}, we parameterize this difference as function of $R$: $H^{(0)}\dfrac{\alpha_s}{2\pi}(-1.3 R+12.0 R^2)$ for the range of $0.2<R<0.6$, which has been included in $H^{(1)}$.

\begin{figure}
\includegraphics[width=0.238\textwidth]{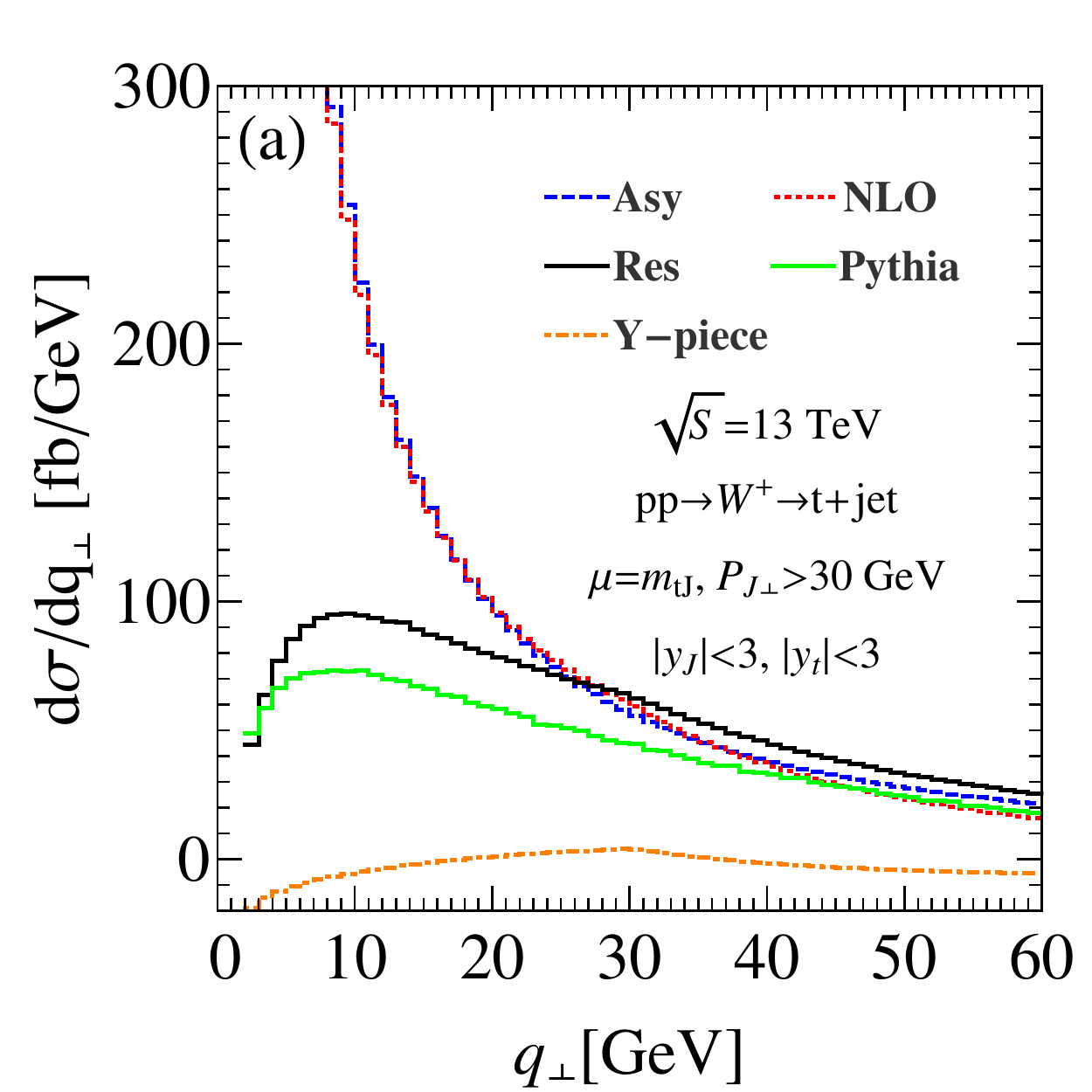}
\includegraphics[width=0.238\textwidth]{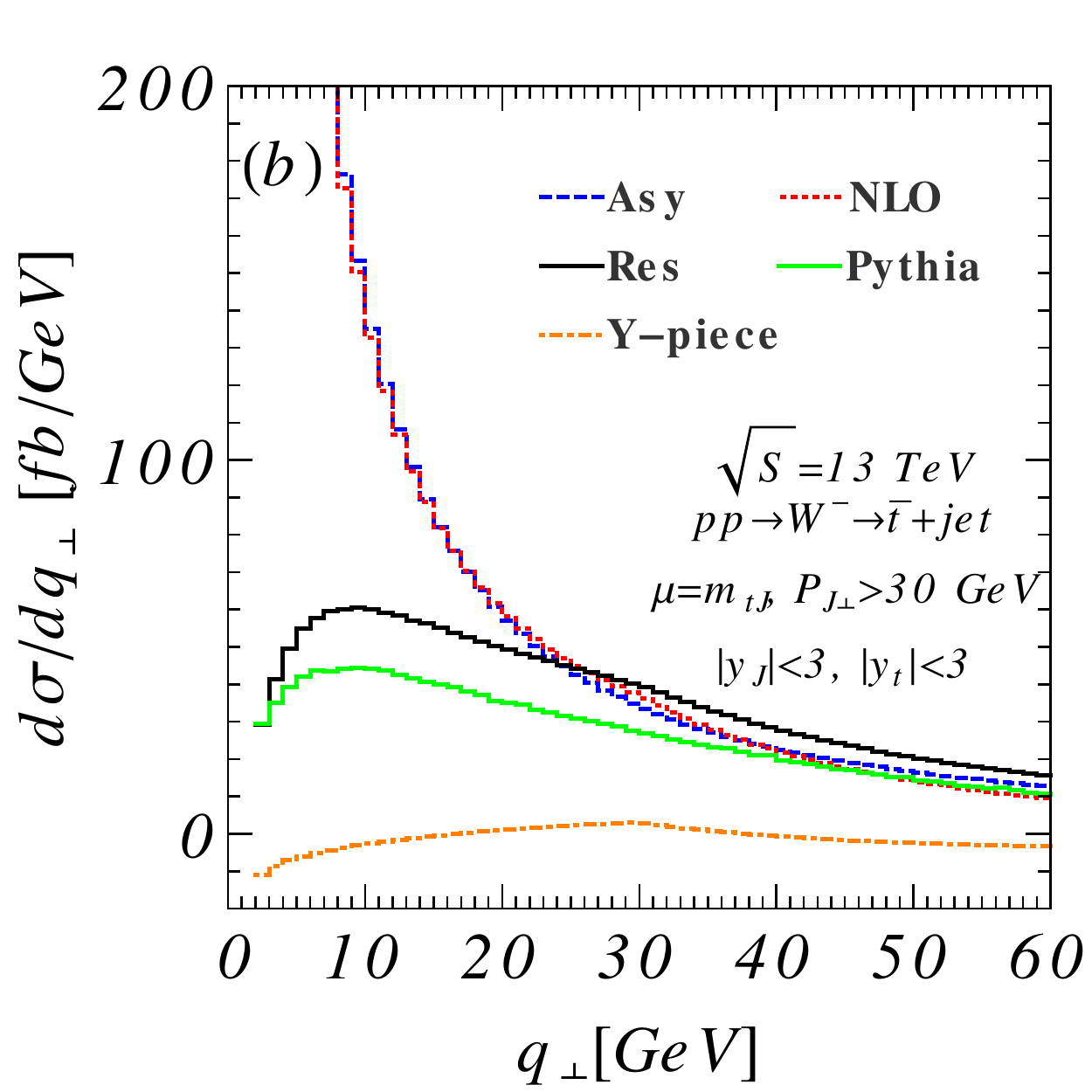}
\caption{ The $q_{\perp}$ distribution from the asymptotic result (blue dashed  line), NLO calculation (red dotted  line), resummation prediction (black solid line), parton shower Monte Carlo  prediction by PYTHIA 8 (green solid line), and $Y$-term (orange dot-dashed line) for the $s$-channel single top quark  (a) and anti-top quark (b)  production at the $\sqrt{S}=13~{\rm TeV}$ LHC with  $|y_J|< 3$,  $|y_t|<3$ and $P_{J\perp}>30~{\rm GeV}$.  The resummation and renormalization scales are choose as $\mu=\mu_{\rm Res}=\mu_{\rm ren}=m_{tJ}=(p_t+p_J)^2$. }
\label{fig:tqt13}
\end{figure}

\begin{figure}
\includegraphics[width=0.238\textwidth]{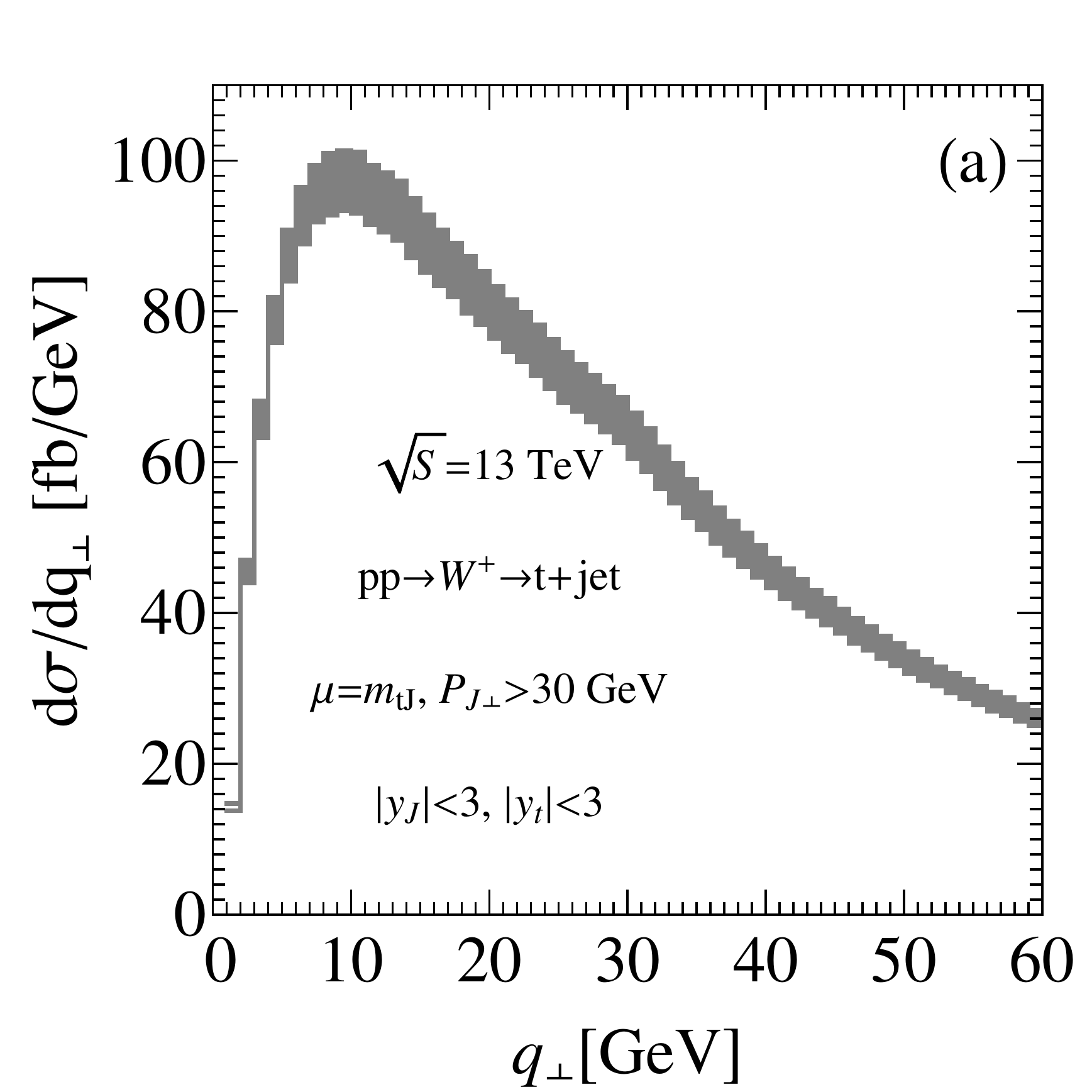}
\includegraphics[width=0.238\textwidth]{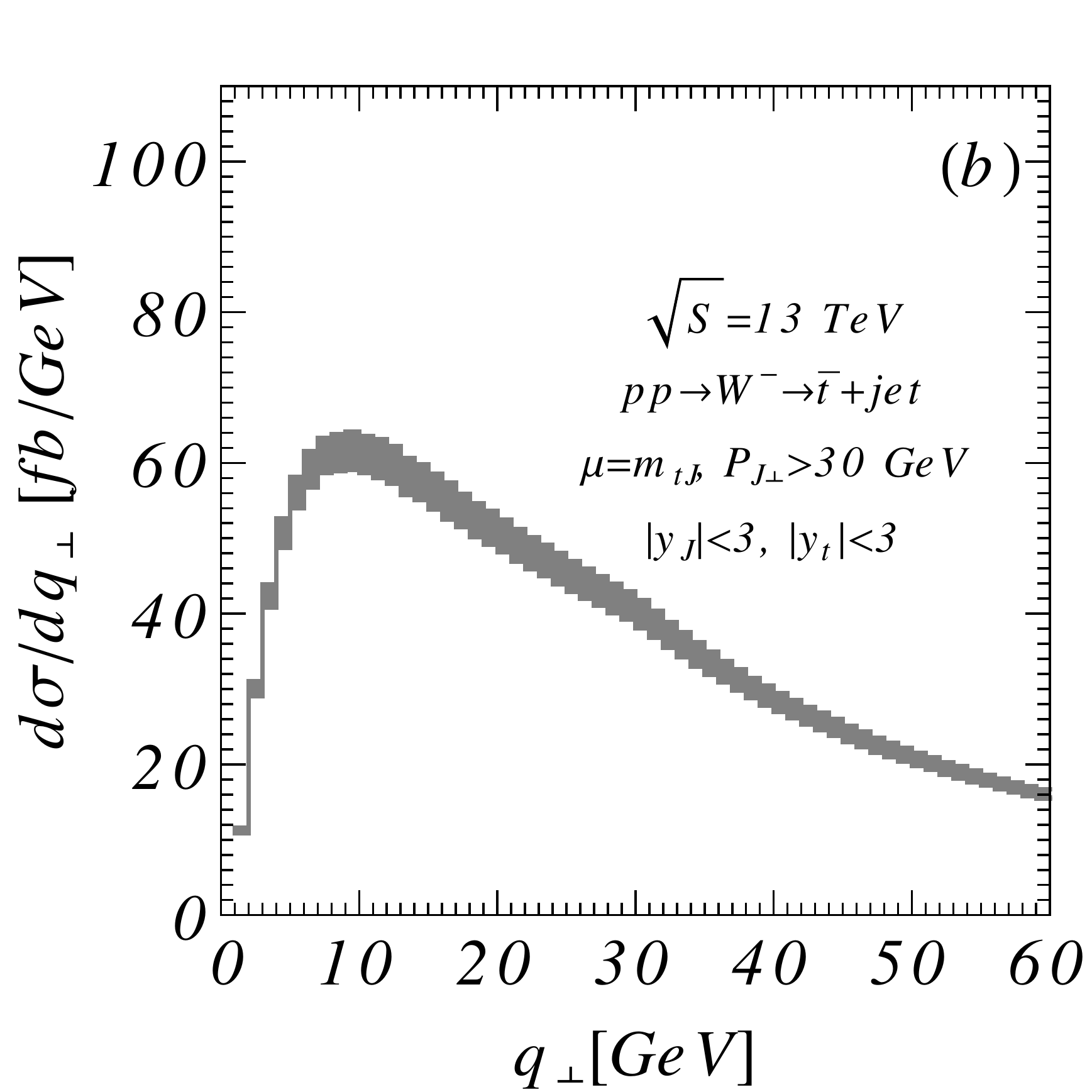}
\caption{ The scale uncertainties for the $s$-channel single top quark (a) and anti-top quark (b)  production at the $\sqrt{S}=13~{\rm TeV}$ LHC with $|y_J|< 3$, $ |y_t|<3$  and $P_{J\perp}>30~{\rm GeV}$. The resummation and renormalization scales are varied from $m_{tJ}/2$ to $2m_{tJ}$. }
\label{fig:scale}
\end{figure}

\begin{figure}
\includegraphics[width=0.238\textwidth]{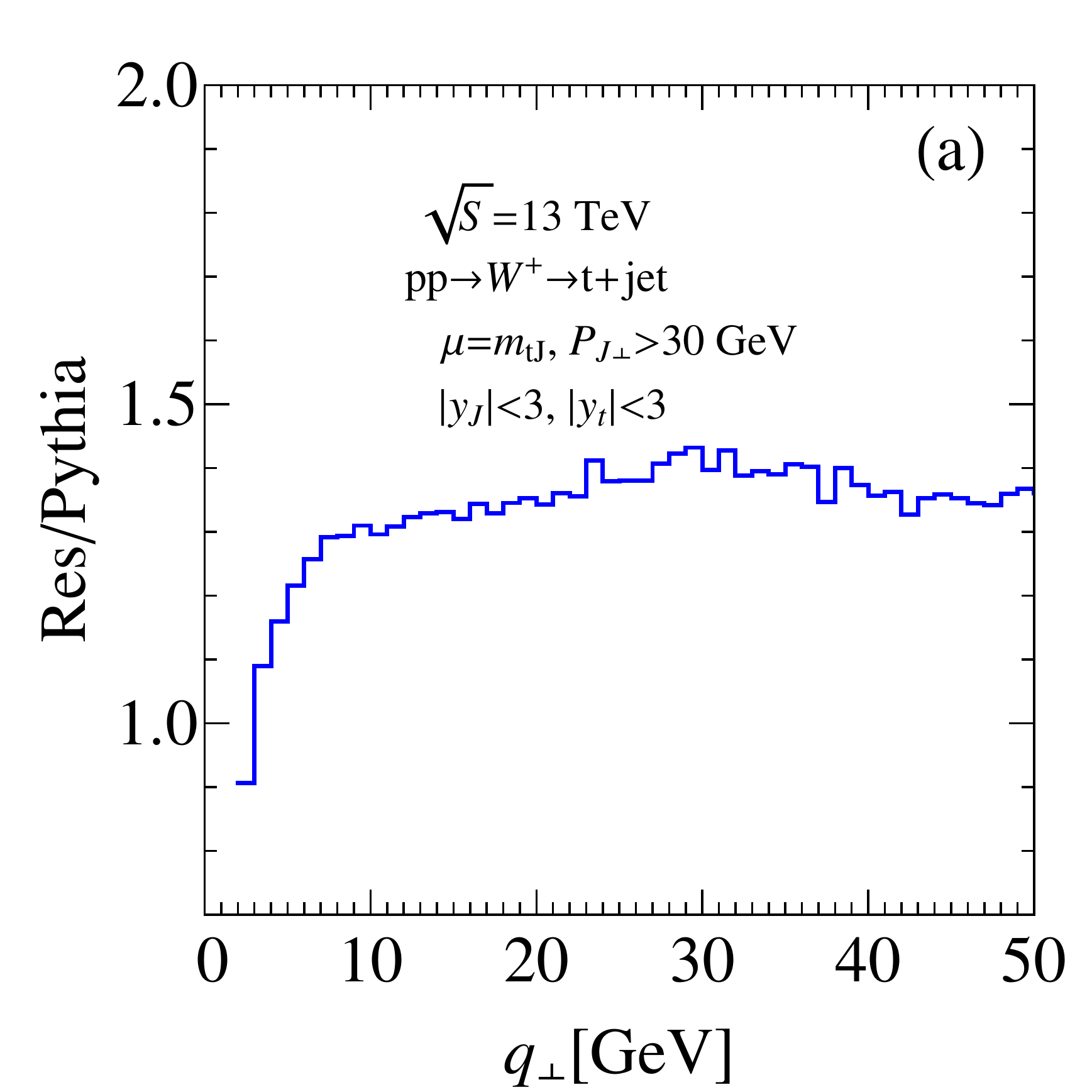}
\includegraphics[width=0.238\textwidth]{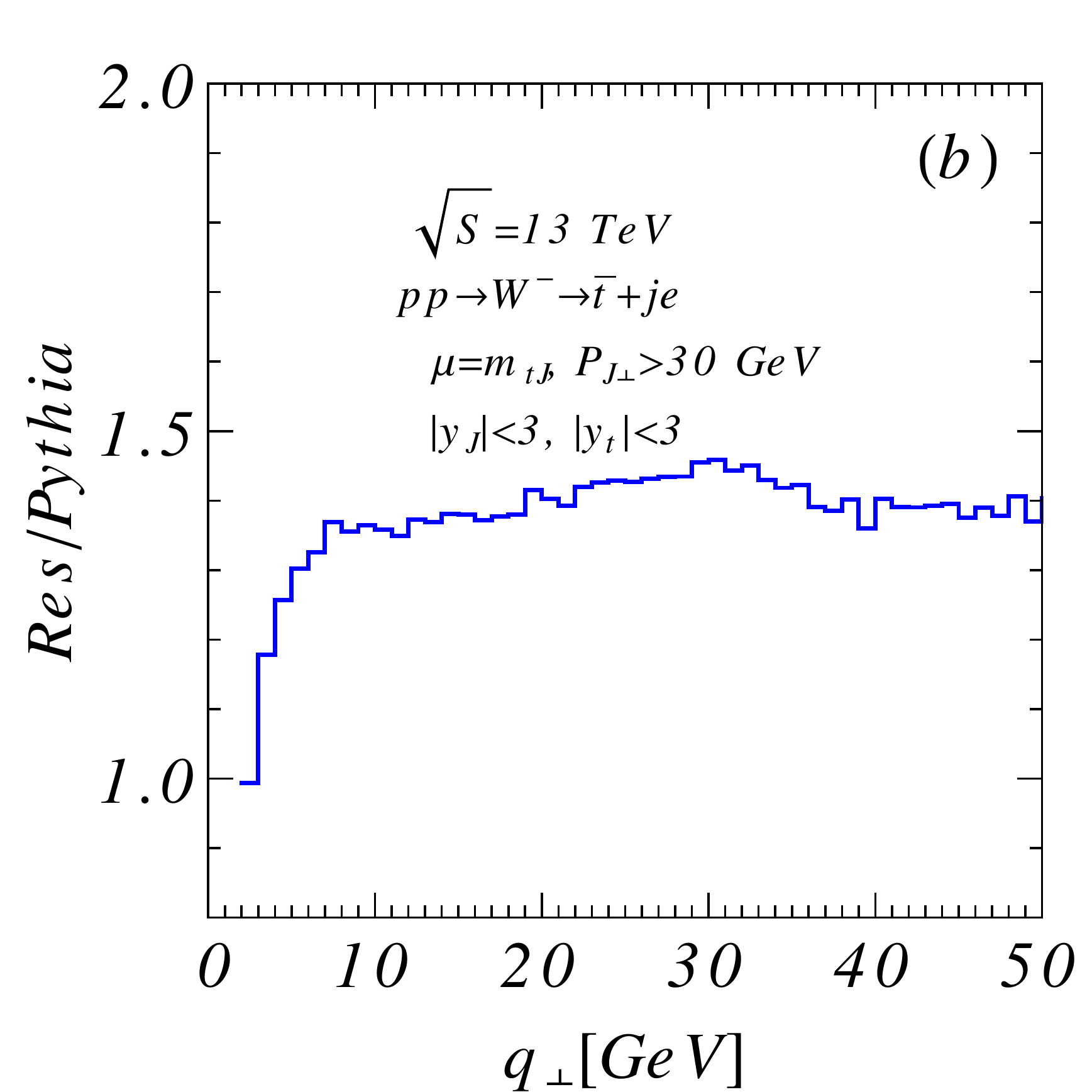}
\caption{ The ratio of the resummation and PYTHIA 8 prediction for the $s$-channel single top quark (a) and  anti-top quark  (b) production at the $\sqrt{S}=13~{\rm TeV}$ LHC with $|y_J|<3, |y_t|<3$ and $P_{J\perp}>30~{\rm GeV}$ . The resummation and renormalization scales are choose as $\mu=\mu_{\rm Res}=\mu_{\rm ren}=m_{tJ}$. }
\label{fig:ratio}
\end{figure}

\begin{figure}
\includegraphics[width=0.238\textwidth]{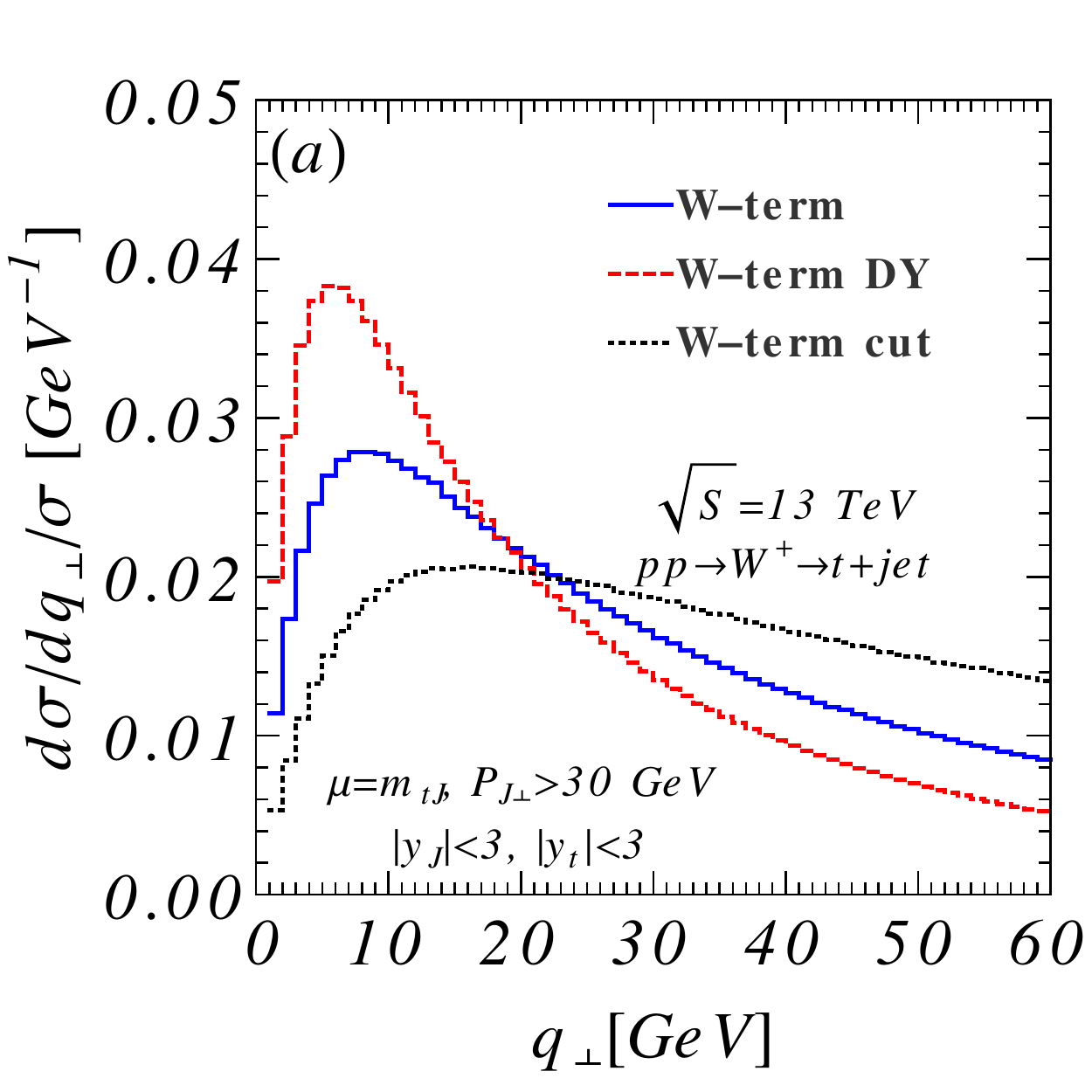}
\includegraphics[width=0.238\textwidth]{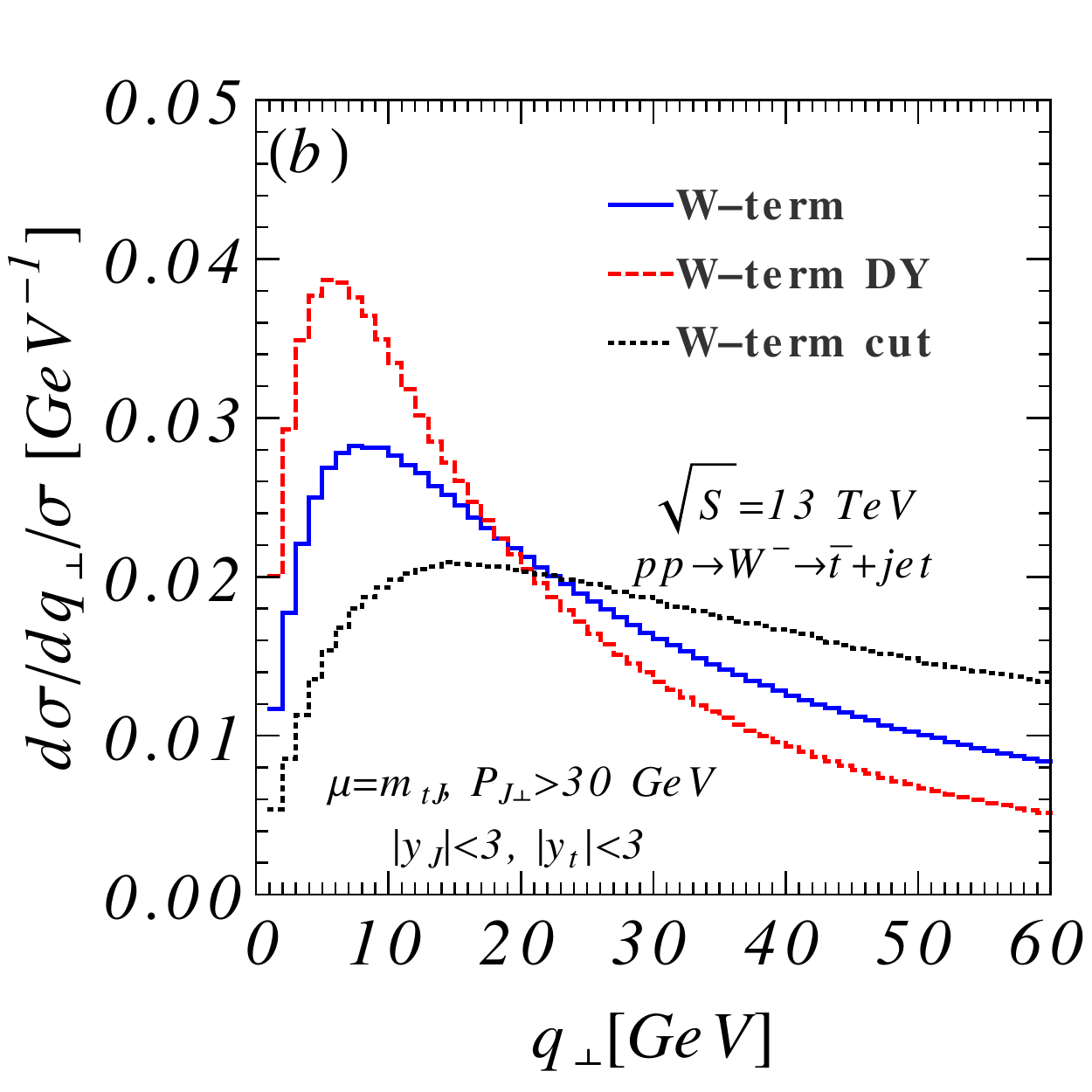}
\caption{The  normalized $W$-piece prediction  for the $s$-channel single top quark (a) and anti-top quark (b)  production at the  $\sqrt{S}=13~{\rm TeV}$ LHC with $|y_J|<3, |y_t|<3$ and $P_{J\perp}>30~{\rm GeV}$. The red-dashed line denotes the $W$-piece prediction  with only Drell-Yan like Sudakov factor, while the  dotted black and  solid blue  lines label the results from the $W$-piece with and without $m_{tJ}>1~{\rm TeV}$ cut. The resummation and renormalization scales are choose as $\mu=\mu_{\rm Res}=\mu_{\rm ren}=m_{tJ}$.}
\label{fig:wpiece}
\end{figure}

Figure~\ref{fig:tqt13} shows  various differential cross sections of the $s$-channel single top quark (a) and anti-top quark (b)  production at the $\sqrt{S}=13~{\rm TeV}$ LHC with  CT14NNLO PDFs
~\cite{Dulat:2015mca}, which were determined in the variable flavor general mass scheme (VFGM) up to five flavors.
Here, we have included the contribution in which bottom quark is one of the incoming partons and taken to be massless in the constituent cross section calculations.   The blue dashed line for asymptotic piece, red dotted line for NLO calculation, black solid line for our resummation prediction, and orange dot-dashed line for the $Y$-term. The asymptotic piece is the fixed-order expansion of Eq.~(\ref{resumy}) up to the $\alpha_s$ order.  In our  resummation calculation, the resummation scale ($\mu_{\rm Res}$) and renormalization ($\mu_{\rm ren}$) scales are taken to be the invariant mass of top quark and jet ($m_{tJ}=(p_t+p_J)^2$). Similarly, the renormalization and factorization scales are also fixed to $m_{tJ}$ in the fixed order calculation. The cone size $R=0.4$ and anti-$k_T$ jet algorithm are used to define the observed jet. The following kinematic cuts are also required in our numerical calculation, $|y_J|<3$, $|y_t|<3$ and $P_{J\perp}>30~{\rm GeV}$. In the same figure, we also compared to  the prediction from the parton shower event generator PYTHIA8 ~\cite{Sjostrand:2007gs} (green solid line), which was calculated at the leading order, with CT14LO PDF.  The uncertainties of the resummation predictions are estimated by varying the scale $\mu_{\rm Res}=\mu_{\rm ren}$ by a factor of two around the central value $m_{tJ}$ , which is shown in  Fig.~\ref{fig:scale}.  
Note that the uncertainty bands could be slight different if we vary the resummation and renormalization
scales independently in the calculation.  In Fig.~\ref{fig:ratio}, we compare the prediction from our resummation calculation to PYTHIA by taking the ratio of their $q_\perp$ differential distributions in Fig.~\ref{fig:tqt13}. 
As shown,  its ratio is not sensitive to $q_\perp$ for either single top (a) or anti-top quark (b) production when $q_\perp>10~{\rm GeV}$, but not for the small $q_\perp$ region. Hence, they predict almost the same shape in $q_\perp$ distribution, while they predict different fiducial total cross section because PYTHIA prediction includes only leading order matrix element and is calculated  with CT14LO PDF.  It would also be interesting to compare our resummation prediction with
that from a parton shower Monte Carlo event generator with NLO matrix element 
which is however beyond the scope of this work.

 In order to estimate the soft gluon radiation effects from the final state, we show various normalized $W$-pieces predictions  in  Fig.~\ref{fig:wpiece}.  The red dashed line denotes the $W$-piece prediction when we only keep the 
 Drell-Yan like Sudakov factor, i.e. the parameters in Eq.~(\ref{su_one})  are changed as $D_1=D_2=0$ and $B_1=-\dfrac{3}{2}C_F\dfrac{\alpha_s}{\pi}$ (labelled as `$W$-term DY' in Fig.~\ref{fig:wpiece}). The dotted black (labelled as `$W$-term cut') and solid blue lines label the results from the $W$-piece including all the Sudakov factor with and without $m_{tJ}>1~{\rm TeV}$ cut, respectively. It shows  that the soft gluon radiation effects from final state are significant in this 
case, especially when we focus on the boosted kinematical phase space region where the term $\left(\ln\dfrac{Q^2-m_t^2}{P_{J\perp}^2R^2}+\ln\dfrac{Q^2-m_t^2}{m_t^2}\right)$  in the Sudakov factor  becomes large. Consequently, the $q_{\perp}$ distribution peaks at a larger value.

\begin{figure}
\includegraphics[width=0.238\textwidth]{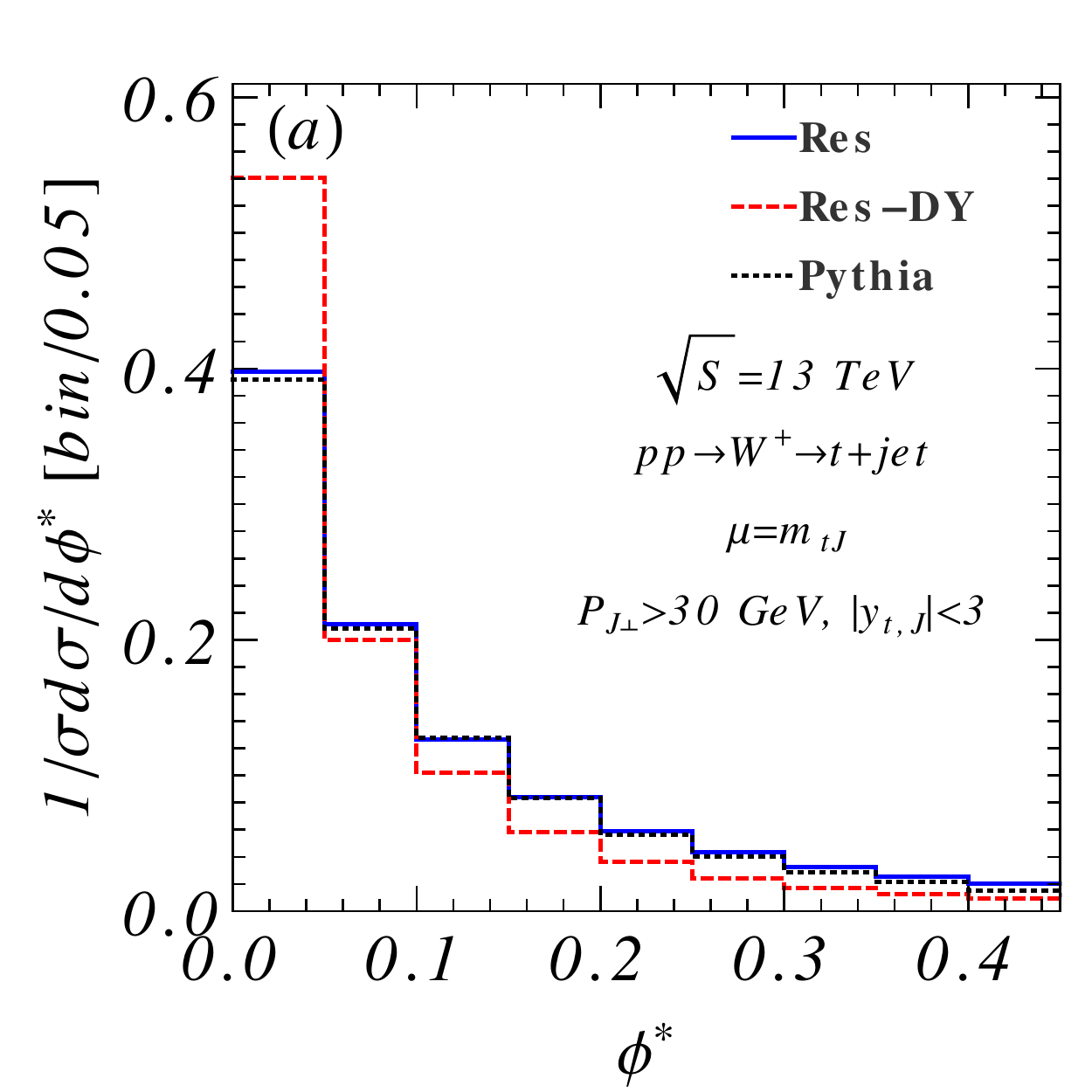}
\includegraphics[width=0.238\textwidth]{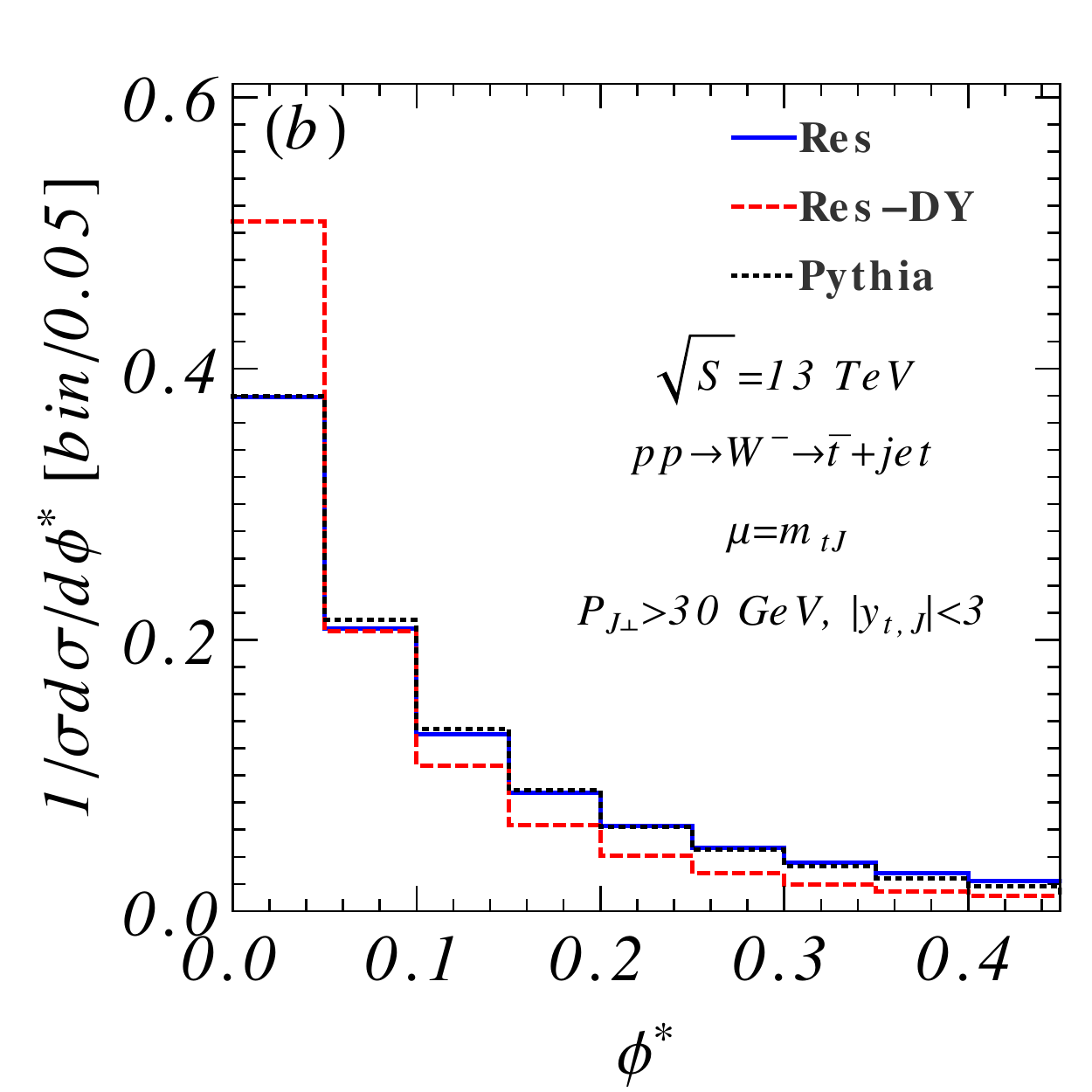}
\caption{The normalized distribution of $\phi^*$ for top quark  (a) and anti-top quark (b)  production at the 13 TeV LHC with $|y_{t,J}|<3$ and $P_{J\perp}>30~{\rm GeV}$. The resummation and renormalization scales are choose as $\mu=\mu_{\rm Res}=\mu_{\rm ren}=m_{tJ}$. The blue solid and red dashed line represent the resummation  prediction with full Sudakov factor and only Drell-Yan  like Sudakov factor, respectively.  The black dotted line labels the prediction from PYTHIA 8.}
\label{fig:phistar}
\end{figure}

Similar to the $t$-channel single top quark production, we can define the $\phi^*$ observable to study the soft gluon radiation effects~\cite{Banfi:2010cf,Cao:2018ntd}. Since the $\phi^*$ only depends on the moving directions (not energies) of the final state jet and top quark, it might  reduce the experimental uncertainties and provide a better measurement for probing the soft gluon radiation effects.  The definition is,
\bea
\phi^*=\tan\left(\frac{\pi-\Delta\phi}{2}\right) \sin \theta^*_\eta,
\eea
where $\Delta\phi$ is the azimuthal angle separation in radians
between the jet and top quark. The angle $\theta^*_\eta$ is defined as,
\bea
\cos\theta^*_\eta=\tanh\left[\frac{\eta_J-\eta_t}{2}\right],
\eea
where $\eta_J$ and $\eta_t$ are the pseudorapidities of the jet and top quark, respectively.
As show in Fig.~\ref{fig:phistar}, the prediction of PYTHIA (black dotted line) and our resummation calculation (blue solid line; labelled as `Res') are consistent with each other. However, if we only keep the Drell-Yan like Sudakov factor in the $W$-term, the $\phi^*$ tends to a smaller value (red dashed line;  labelled as `Res-DY' in Fig.~\ref{fig:phistar}).  It could be understood from the $W$-piece prediction  in Fig.~\ref{fig:wpiece}, where the Sudakov factor from the final state soft gluon radiation would push the $q_\perp$ distribution to peak at  a larger 
$q_\perp$ value.  Because  large $q_\perp$ value corresponds to the large $\phi^*$ value, the final state soft gluon radiation would push the average $\phi^*$ value  to a larger value as compared to the Drell-Yan like Sudakov factor case.

\section{Phenomenology of top-flavor $W^\prime$ production}

\begin{figure}
\includegraphics[width=0.24\textwidth]{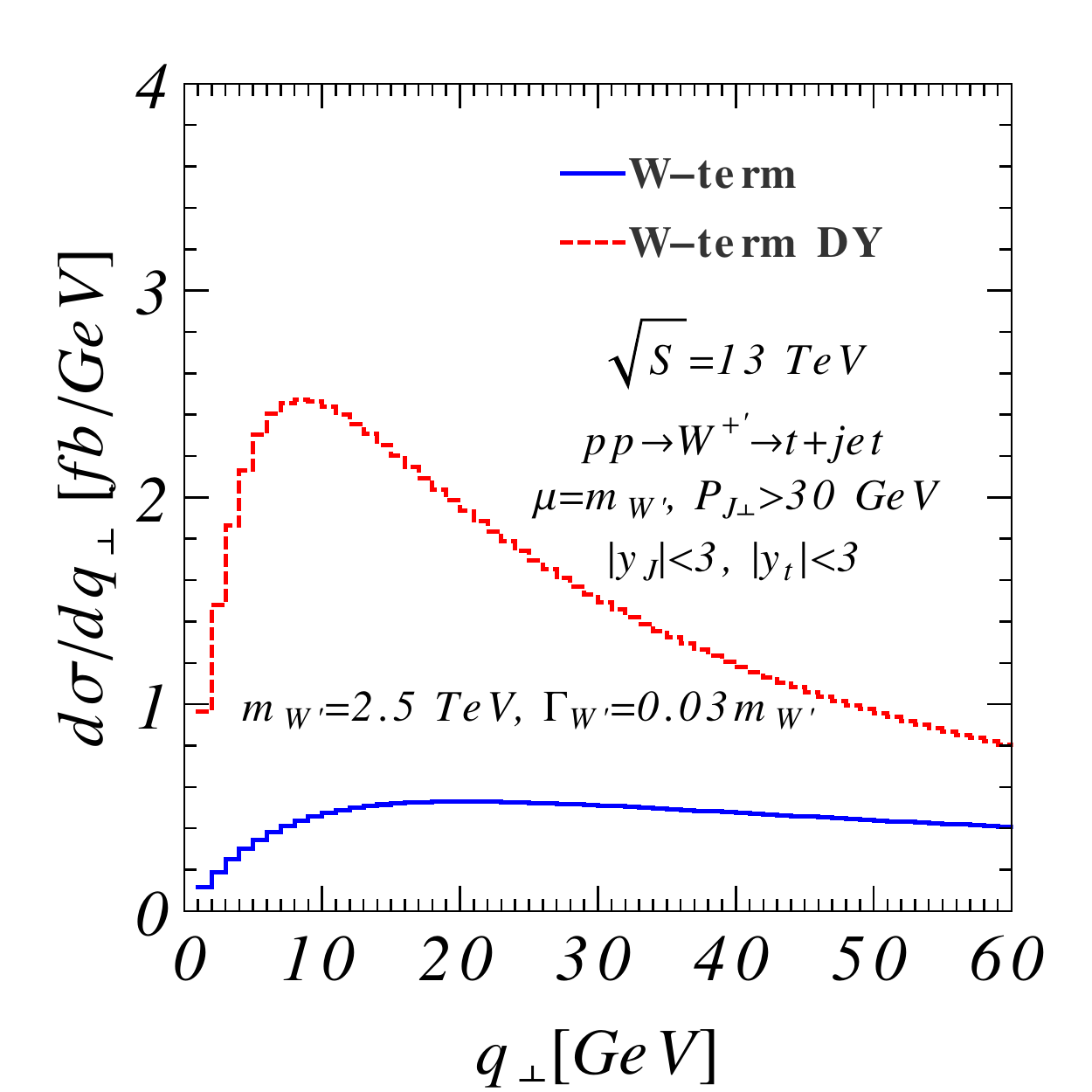}
\caption{The  prediction  for  $pp\to W^{\prime+}\to t+jet$ at the  $\sqrt{S}=13~{\rm TeV}$ LHC with $|y_J|<3, |y_t|<3$ and $P_{J\perp}>30~{\rm GeV}$. The red-dashed line denotes the $W$-piece prediction with only Drell-Yan like Sudakov factor, while the  solid blue  line labels  the result from the $W$-piece with full Sudakov factor. The resummation and renormalization scales are choose as $\mu=\mu_{\rm Res}=\mu_{\rm ren}=m_{W^\prime}$.}
\label{fig:wprime1}
\end{figure}

In many new physics models,  extra heavy particles would favor decay into top quark and jet, e.g.  the $W^\prime$ in top-flavor models~\cite{Li:1981nk,Malkawi:1996fs,He:1999vp,Hsieh:2010zr,Cao:2012ng} or a  charged Higgs $H^+$ in the general two Higgs doublet models~\cite{He:1998ie}. When $m_{W^\prime/H^+} \gg m_t$, the decayed  top quark and jet are highly boosted. As we discussed in the last section, the Sudakov enhancement  from the final state soft gluon radiation would become more important  when top quark and jet are highly boosted.   In this section,  we  use the Sequential Standard  Model (SSM) $W^\prime$  as  an example to discuss the final state  soft gluon radiation effects on the kinematical distribution of the $W^\prime$. Our results can easily be extended to the general $W^\prime$ new physics models. The effective lagrangian related to our study is,
\bea
\mathcal{L}=\dfrac{g}{\sqrt{2}}V_{ij}\bar{q}_i\gamma_\mu\dfrac{1-\gamma_5}{2}q_jW^{\prime\mu}.
\eea
The null result in  the search of $W^\prime$ via the single top quark channel at the 13 TeV LHC by ATLAS and CMS collaborations  impose a strong bound on $m_{W^\prime}$, which should be larger than 2-3 TeV~\cite{Sirunyan:2017vkm,Aaboud:2018juj}. In this work, we assume $m_{W^\prime}=2.5~{\rm TeV}$ and $\Gamma_{W^\prime}=0.03 m_{W^\prime}=75~{\rm GeV}$ as our benchmark point to study the soft gluon radiation effects on the transverse momentum distribution of $W^\prime$ . 

The $W$-term predictions with full Sudakov factor (blue solid line) and Drell-Yan like Sudakov factor (red-dashed line) at the 13 TeV LHC are shown in Fig.~\ref{fig:wprime1}. 
Clearly, the soft gluon radiation effects from the final state jet and top quark are important for the search of heavy resonance states in $s$-channel single-top processes.

\section{Conclusion}
In this work,  we studied the $q_\perp$ resummation effects for the $s$-channel single top quark production at the LHC based on the TMD factorization theorem.  The  large logarithm $\ln(Q^2/q_{\perp}^2)$ was resummed by renormalization group evolution at NLL accuracy. 
We also calculated the NLO total cross section derived from the resummation framework, while yields a slightly different result
 from the MCFM prediction due to the usage of narrow jet approximation in our resummation calculation. To ensure the correct NLO total cross section, we have added an additional term proportional to $H^{(0)}$ to account for the above difference in our resummation calculation. A detailed comparison between our theory calculation and PYTHIA 8 prediction was also discussed.  We find  that  both  the total transverse momentum ($q_\perp$) and $\phi^*$ distributions predicted by  our theory calculation agree well with the PYTHIA 8  prediction.
Furthermore,  the soft gluon radiation effects from the final state would change the shape of $q_\perp$ or $\phi^*$ distribution significantly, especially when the top quark and jet are highly boosted.  Finally, we discussed the  soft gluon radiation effects for the  production of $W^\prime$ or $H^+$ boson which subsequently decays into a pair of top quark and jet.

\begin{acknowledgments}
This work was supported  by the U.S. National
Science Foundation under Grant No. PHY-1719914.
C.-P. Yuan is also grateful for the support from 
the Wu-Ki Tung endowed chair in particle physics.
\end{acknowledgments}

\bibliographystyle{apsrev}
\bibliography{reference}

\end{document}